\documentclass[prl,aps,twocolumn,superscriptaddress,showpacs,floatfix]{revtex4}
\usepackage{graphicx}
\usepackage{amsmath, amssymb}

\begin{document}

\newcommand{\Do}{D_o}   
\newcommand{\tauR}{\tau_R}   
\newcommand{\tauo}{\tau_o}   

\title{Polymer Translocation in Crowded Environments}
\author{Ajay Gopinathan}
\address{School of Natural Sciences, University of California,
Merced, CA 95344 ,U.S.A.}
\author{Yong Woon Kim}
\address{Department of Physics and Materials Research Laboratory,
University of California, Santa Barbara, CA 93106 ,U.S.A.}

\begin{abstract}
We study the effect of the crowded nature of the cellular
cytoplasm on the translocation of a polymer through a pore in a
membrane. By systematically treating the entropic penalty due to
crowding, we show that the translocation dynamics are
significantly altered, leading to novel scaling behaviors of the
translocation time in terms of chain length. We also observe new
and qualitatively different translocation regimes depending upon
the extent of crowding, transmembrane chemical potential
asymmetry, and polymer length.

\end{abstract}


\maketitle

Transport of a variety of biopolymers across a dividing membrane
is a fundamentally important process in biological systems~\cite{lodish}.
Examples include transfer of proteins across
cellular membranes or endoplasmic reticulum~\cite{schatz}, gene
swapping through bacterial pili~\cite{miller} and RNA transport
through nuclear pore complexes~\cite{npc}. Technological
applications include gene delivery~\cite{gene} and DNA sequencing~\cite{meller}.
A considerable amount of theoretical work has
focused on both the basic physics~\cite{park1,muthu1,mehran,oster,roya}
underlying the translocation process and on how
details such as polymer-pore interactions~\cite{kolomeisky},
intrinsic polymer structure~\cite{ulrich} and
confinement~\cite{muthu2,park2} affect the dynamics of the
process. One important aspect that has received very little attention is the
effect of crowding on the translocation dynamics. Crowding due to
macromolecular aggregates and other inclusions in the cellular
cytoplasm can be as high as 50\% by volume~\cite{crowding1}
and is known to have considerable influence on reaction rates,
protein folding rates and equilibria {\it in vivo}~\cite{minton,zhou}.
A polymer threading its way through such a crowded
environment is subject to a large entropic penalty which should
dramatically affect the translocation dynamics. In this Letter we
present the first systematic study of
polymer translocation in terms of crowding.
We show that the free energy penalty due to crowding has terms
that scale sublinearly with the polymer length.
The presence of non-linear terms in the free energy leads to
qualitatively different translocation dynamics including novel
power law scalings of the translocation time with polymer length
as well as situations where the translocation time is nearly {\it
independent} of the polymer length over several orders of magnitude.

\begin{figure}[b]
\begin{center}
\resizebox{7 cm}{!}{\includegraphics{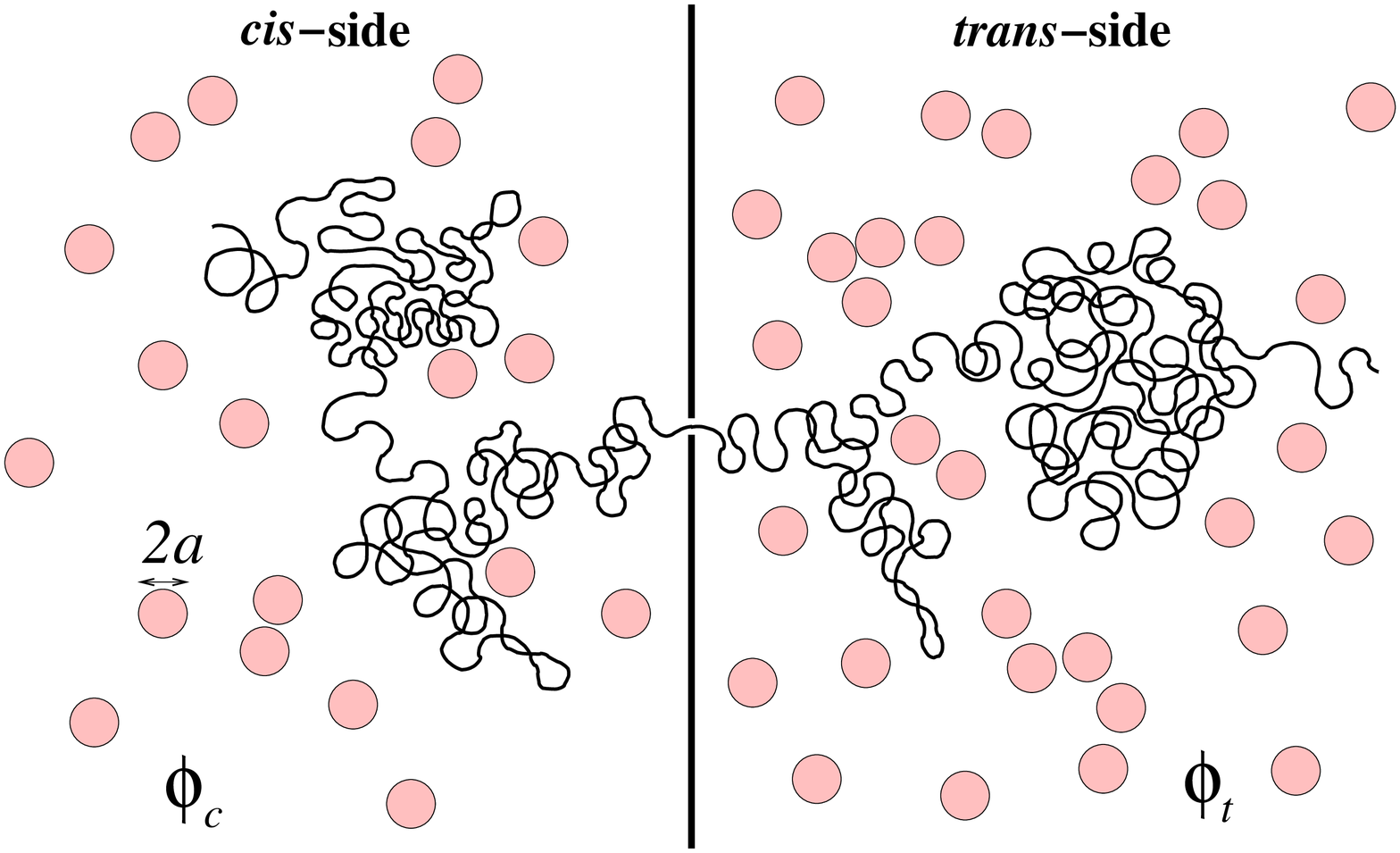}}
\end{center}
\caption{Schematic illustration of the translocation process
of a polymer in the presence of crowding }
\label{fig:schematic}
\end{figure}

We consider 
a Gaussian polymer of length $N$ (in units of the Kuhn length $b$)
threading itself through a pore in a dividing membrane from
{\it cis} (left) side to {\it trans} (right) side,
as illustrated in Fig.~\ref{fig:schematic}.
The pore is assumed to be small enough that
it allows only one monomer to pass through at a time, with an
effective diffusion constant $D_{P}$. Crowding is modeled by
randomly distributed spherical obstacles, sterically interacting with
the polymer, of radius $a$ and diffusion
constant $\Do$ at a volume fraction $\phi_c$($\phi_t$) on the
{\it cis} ({\it trans}) side. There could also be an
excess chemical potential difference for monomers between
{\it cis} and {\it trans} side, $\Delta \mu$.
We now assume that we can treat the process quasi-statically with the
polymer segments on both sides of the membrane being in
equilibrium at all times. The validity of the assumption depends
on the relative magnitudes of the three time scales in the
problem: the total translocation time $\tau \sim ( b^2/D_{P})
\tilde{\tau}$, the polymer relaxation time $\tauR \sim (\eta
b^3/k_BT ) N^2$~\cite{doi}, and the timescale set by obstacle motion
$\tauo \sim \phi^{-2/3}/\Do$. Here $\tilde{\tau}$
is a dimensionless function that characterizes the translocation
process and $\eta$ is the medium's viscosity. Assuming equilibrium
statistics for the polymer segments necessarily requires that
$\tauR\ll\tau$~\cite{park1}. In the presence of obstacles, this
assumption remains valid in two different regimes, $\tauo \ll
\tauR \ll \tau$ and $\tauR \ll \tau \ll \tauo$. The first regime
corresponds to the situation where the obstacles diffuse fast
enough that the ``polymer segment + obstacles'' system can be
assumed to be at equilibrium (dynamic obstacles).
In the second regime, the obstacles
are essentially immobile on the translocation timescale and the
polymer segment achieves equilibrium statistics in this static
obstacle environment. It should be noted that if the pore friction
is not high enough (i.e. $\tauR \simeq \tau$)~\cite{mehran}
or if $\tauR \leq \tauo \leq \tau$ the quasi-static
assumption breaks down leading to anomalous dynamics. In the
regimes where the assumption remains valid we have a well defined
free energy barrier whose form is governed by the polymer
statistics, presence of the membrane, chemical potential gradient
and the presence of crowding. Since the contributions to the free
energy from factors other than crowding have been worked out
before~\cite{park1,muthu1}, we focus on the entropic penalty that
arises from crowding.

Polymer configurations in the presence of static obstacles
correspond to Brownian walk trajectories with a diffusion constant
$b^2/6$ that have survived to a time $t=n$ with the obstacles
playing the role of traps. The fraction of allowed polymer
configurations is therefore identical to the survival probability
of such a Brownian walker, leading to the free energy expression
for the entropic penalty in units of $k_BT$ as
\begin{equation}
 F^s_{cr}(n) = - \mbox{log } S(n) ,
 \label{penalty}
\end{equation}
where $S(t)$ is the survival probability of an appropriate
Brownian walker at time $t$~\cite{zhou,machta}. For short times,
$S(t)$ is given by the Smoluchowski solution
$S(t) \sim \exp{(-\kappa t)}$~\cite{smol1, smol2}, while for long times it is
dominated by walkers trapped in large void regions giving rise to
the stretched exponential Donsker-Varadhan (DV) solution
$S(t) \sim \exp{(-\lambda t^{3/5})}$~\cite{dv1,dv2}, where $\kappa$ and
$\lambda$ are constants that depend on trap radius, trap density
and geometry. Using  Eq.~(\ref{penalty}) and the exact solutions for
$S(t)$~\cite{smol1,dv1}, we can explicitly compute the free energy
penalty for a chain of length $n$ due to static obstacles at
volume fraction $\phi$, yielding
\begin{eqnarray}
F^s_{cr}(\phi,n)&=&\frac{1}{2} \left(\frac{b}{a}\right)^2 \phi
\left(n + \sqrt{\frac{24 a^2 n}{\pi b^2}}  \right)
~\mbox{for}~~ n\ll N_{\times} \nonumber \\
&=&2.6 \left(\frac{b}{a}\right)^{6/5}
\phi^{2/5} n^{3/5} ~~~ \mbox{for}~~ n \gg N_{\times}  ,
\label{fcrstat}
\end{eqnarray}
where $N_{\times} \sim (a/b)^2 \phi^{-3/2}$ represents the
crossover polymer length
from the Smoluchowski regime to the DV regime~\cite{dvcross}.
For the dynamic obstacle case, the ``polymer + obstacles'' system comes to
equilibrium. The presence of the obstacles gives rise to a
depletion induced attractive interaction between chain monomers
that can lead to a collapsed polymer phase similar to that induced
by poor solvent conditions~\cite{degennes}.
Simulations~\cite{frenkel} and analytical work
\cite{thirumalai,paul} have shown that hard spheres can cause
polymers to collapse if the sphere density is high enough. For the
range of $\phi$ ($0.1<\phi<0.5$) that we are interested in, the
collapse occurs whenever $a \ge 2b$. For our purposes it is then
safe to assume that our Gaussian polymer is in a collapsed ``dense
globule'' state with a volume $Nb^3$. The free energy penalty is
then given by the sum of the confinement entropic penalty ($=
\pi^2 R_G^2/6 R^2$ where $R \sim N^{1/3}b$ is the confining
radius) and the work done to create a cavity of volume $Nb^3$ that
is devoid of obstacles, which is known exactly from scaled
particle theory \cite{SPT}. The resulting
free energy penalty for the chain due to dynamic obstacles reads as
%
\begin{figure}
\begin{center}
\resizebox{8 cm}{!}{\includegraphics{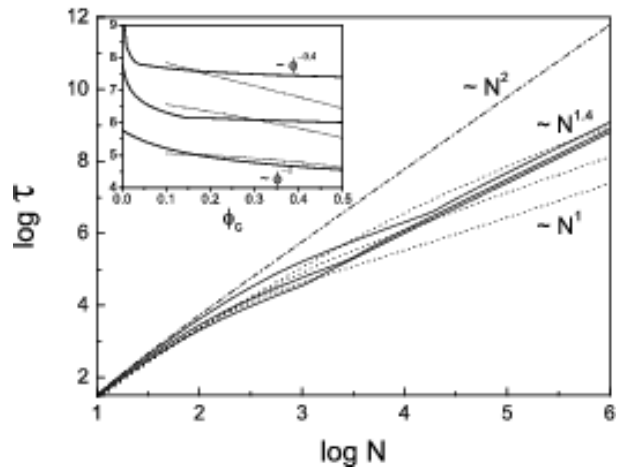}}
\end{center}
\caption{ Translocation time $\tau$ (in units of $b^2/D_P$)
of a polymer of length $N$ releasing out of the crowded {\it cis} side
with static obstacles (solid lines: $\phi_c=0.1, 0.3, 0.5$ in descending order),
dynamic obstacles (dotted lines: for the same values of $\phi_c$),
and no crowding (dashed-dotted line: $\phi_c=0$).
Inset: $\tau$ vs. obstacle volume concentration $\phi_c$
for static (solid lines: $N=10^5, 10^4, 10^3$ in descending order) and
dynamic obstacles (dotted lines: for the same values of $N$).
Note $\phi_t=\Delta \mu=0$ here and the value of $b/a=0.3$ is used throughout
the paper.}
\label{fig:escape}
\end{figure}
\begin{eqnarray}
 F^d_{cr}(\phi,n) &\sim& 
 \frac{3b^3(1 +\phi+\phi^2) \phi n}{4\pi a^3(1-\phi)^3}
  - \frac{9}{2}
  \left(\frac{3}{4\pi}\right)^{\frac{2}{3}}
 \frac{ b^2\phi^2(1+\phi)n^{\frac{2}{3}}}{ a^2(1-\phi)^3} \nonumber \\
 & +& \left( \frac{\pi^2}{6}+9\left(\frac{3}{4\pi}\right)^{1/3}
 \frac{b \phi^3}{a (1-\phi)^3} \right) n^{1/3} .
 \label{fcrdyn}
\end{eqnarray}
The general expression for total free energy, taking into account
the presence of the dividing wall and chemical potential difference,
of a chain with $n$ monomers on the {\it trans} side
and $N-n$ on the {\it cis} side is then given by
\begin{eqnarray}
 F^s_{tot}(n) &=& F^s_{cr}(\phi_c, N-n)+ F^s_{cr}(\phi_t, n) \nonumber \\
 &+&\frac{1}{2}\ln[n(N-n)] + n \Delta \mu  
\label{fstat}
\end{eqnarray}
for the static obstacle case and
\begin{equation}
 F^d_{tot}(n) = F^d_{cr}(\phi_c, N-n)+ F^d_{cr}(\phi_t, n) +n \Delta \mu
\label{fdyn}
\end{equation}
for the dynamic obstacle case. It is to be noted that the
logarithmic term in Eq.(\ref{fstat}),
resulting from reduced chain configurations confined in a half-space,
does not appear in Eq.(\ref{fdyn})
because $F^d_{cr}$ already takes into account the entropic penalty
associated with confining the chain to a dense globule.  The
translocation process can now be described by diffusion along the
translocation coordinate $n$ in the presence of a
well-defined free energy barrier, $F^{d(s)}_{tot}(n)$. The dynamics
of this process is governed by a Fokker-Planck equation which then
allows one to compute the translocation time
(mean first passage time for the chain to diffuse across the pore~\cite{park1})
as
\begin{eqnarray}
 \tau_{d(s)} = \frac{b^2}{D_{P}} \int_{1}^{N-1} d n~ e^{F^{d(s)}_{tot}(n)}
 \int_{1}^{n} d n'~ e^{-F^{d(s)}_{tot}(n')} .
\label{tauexp}
\end{eqnarray}
Since we have nonlinear terms in the free energy we first consider
the behavior of $\tau$ for a general free energy functional with a
power law scaling, e.g., $F(n) \sim n^{\alpha}$. Eq.~(\ref{tauexp})
and saddle point approximations (in the large $N$ limit) to do
the integrals yield
 \begin{eqnarray}
\tau &\sim& N^{2-\alpha} ~~~~~~\mbox{for}~~~F(n) \sim (N-n)^{\alpha} \nonumber \\
\tau &\sim& \exp(N^{\alpha}) ~~\mbox{for}~~~F(n) \sim n^{\alpha}
\label{power}
\end{eqnarray}
for translocation out of and into a crowded half-space,
respectively. Thus we anticipate new exponents characterizing
the scaling properties of the mean passage time with the number of monomers
in the crowded medium.

We now consider, in detail, the impact of crowding on translocation dynamics
in some physically interesting situations. The first example is that of a polymer
escaping from a crowded environment, i.e., where the {\it cis} side is crowded and
the {\it trans} side is not ($\phi_t=0$).
In the static case, the dominant contribution to the free energy has
the form $\phi_c^{2/5} (N-n)^{3/5}$ for a long chain
and $\phi_c (N-n)$ for a short chain (see Eq.~(\ref{fcrstat})).
The translocation times obtained by saddle point approximation
scale as $\phi_c^{-2/5} N^{7/5}$ and $\phi_c^{-1} N$, respectively.
Exact numerical evaluations of Eq.~(\ref{tauexp}) confirm
the predicted power law scalings with both $N$ (Fig.~\ref{fig:escape})
and $\phi_c$ (inset in Fig.~\ref{fig:escape}).
In contrast to the dynamic case where the leading order behavior simply
corresponds to having an effective ``osmotic pressure'' from obstacles,
the driving force for translocation
in randomly distributed immobile obstacles is weakened by the existence of rare,
large voids on the {\it cis} side that sufficiently long polymers can explore.
This is the physical origin for the novel exponents describing the scaling
of $\tau$ with respect to both $N$ and $\phi_c$
for a long enough chain in the static case.

\begin{figure}
\begin{center}
\resizebox{8.7 cm}{!}{\includegraphics{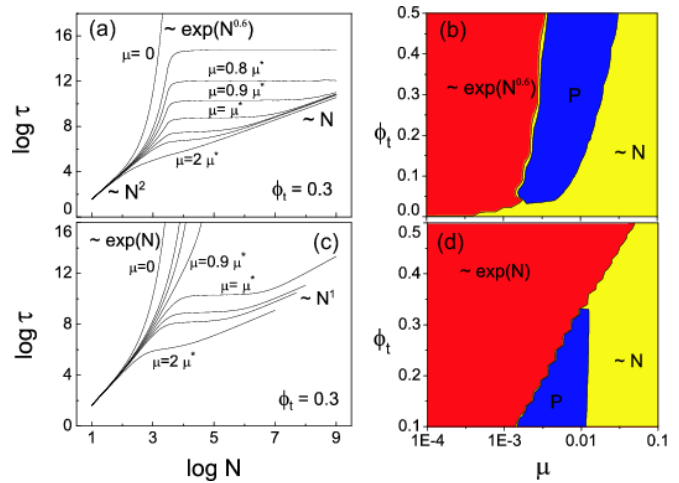}}
\end{center}
\caption{
Translocation time $\tau$ (in units of $b^2/D_P$) vs. $N$ for (a) static and (c) dynamic
obstacles on the {\it trans} side ($\phi_t=0.3, \phi_c=0$) for different
values of chemical potential gradient, showing the asymptotic power
law scalings as well as the plateau regime at intermediate length scales.
Note $\mu^*$ denotes the value of the prefactor of the $n$-linear term of $F_{cr}$
in Eq.~(\ref{fcrstat}) and (\ref{fcrdyn}), respectively.
Different scaling  behaviors (exponential, plateau, power law) of
$\tau$ in the $(\phi_t,\mu)$ phase space for (b) static and (d) dynamic obstacles
at a fixed $N=10^5$.
}
\label{fig:chemical}
\end{figure}

Another situation of significant practical interest is
translocation into a crowded environment driven by a chemical
potential gradient. Here we take $\phi_c=0$ and $\Delta\mu=-\mu$
that favors translocation into the crowded {\it trans} side.
For the static obstacle case in the large $N$ limit,
the linear chemical potential term always dominates
for any non-zero value of $\mu$.
Thus $\tau\sim N$ in this limit,
while, for $\mu=0$, $\tau \sim \exp{(N^{3/5})}$ (from Eq.~(\ref{power})).
The situation is similar for the dynamic case but only if $\mu$ exceeds a critical
threshold $\mu^*$ that is sufficient to overcome the osmotic pressure term
($\mu^*$ is here defined by the value of the prefactor of the $n$-linear
osmotic pressure term in $F_{cr}$).
For $\mu < \mu^*$, $\tau\sim \exp(N)$ for a long chain
when obstacles are mobile.
Fig.~\ref{fig:chemical}(a) and (c) clearly show these distinct limiting behaviors.
The plots also reveal a
striking phenomenon that seems to occur at intermediate length
scales. Depending on the parameter values, there appear to be
regimes spanning several orders of magnitude in polymer length,
where the translocation time is {\it nearly} independent of $N$.
The reason for this can be understood by considering the form of
the relevant free energy profile as a function of the
translocation coordinate $n$. The situation is particularly simple
for the static obstacle case where there is a competition between
a  linear term ($\sim n$) due to the chemical potential and a
sublinear term ($\sim n^{3/5}$) that comes from the crowding for a long chain. This
gives rise to a free energy barrier whose height and position are nearly independent
of the total length $N$ followed by a ``downward slope'' all the
way to $n=N$. The time taken to surmount this barrier, which is
independent of $N$, is rate-limiting and hence effectively the
total translocation time for polymer lengths shorter than the
value of $N$ at which the time taken to traverse the downward
slope becomes comparable. At this point the scaling crosses over
to being linear in $N$. The situation for the dynamic obstacle
case is similar except that the presence of two sublinear terms
``softens'' the plateau because barrier height and position are
no longer independent of $N$. Figs.~\ref{fig:chemical}(b) and (d) show
the different translocation time scaling regimes in the $(\phi_t,
\mu )$ phase space at a fixed $N$. As one might expect, large values
of $\mu$ imply the dominance of the chemical potential, leading to
$\tau\sim N$, while low values give exponential behavior.
The disappearance of the plateau regime for the dynamic
obstacles at high concentrations 
is because the presence of
two sublinear terms becomes more apparent at higher $\phi_t$.

\begin{figure}
\begin{center}
\resizebox{8.7 cm}{!}{\includegraphics{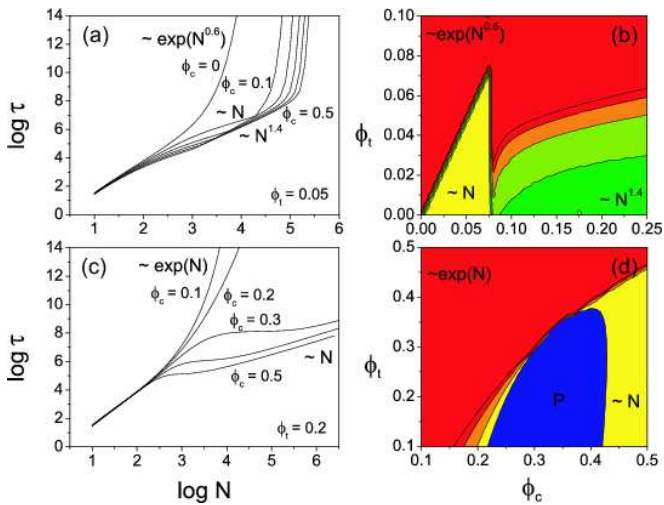}}
\end{center}
\caption{
Translocation time $\tau$ (in units of $b^2/D_P$) as a function of chain length $N$
for (a) static($\phi_t=0.05$) and (c) dynamic obstacles($\phi_t=0.2$)
for different values of $\phi_c$.
Different scaling behaviors (exponential, plateau, power law)
of $\tau$ in the $(\phi_t,\phi_c)$ phase space
for (b) static and (d) dynamic obstacles at a fixed $N=3\times 10^4$.
No chemical potential gradient is applied ($\Delta \mu=0$).
}
\label{fig:both}
\end{figure}

As a final example we consider the polymer translocation
when both sides are crowded and $\Delta \mu=0$.
For the dynamic obstacles, qualitative picture is
similar to that with
{\it trans} side crowding and a chemical potential gradient (compare
Fig.~\ref{fig:chemical}(c,d) with Fig.~\ref{fig:both}(c,d)).
It is to be noted however that here the presence of
sublinear free energy contributions from both sides of the
membrane smears out the plateau even more. In the static case
however, the absence of a linear term in the long polymer limit
gives rise to qualitatively different results. As seen in Fig.~\ref{fig:both}(a),
even a minute amount of $\phi_t$
leads to exponential translocation times at large polymer lengths
despite substantial crowding on the {\it cis} side.
This counter-intuitive behavior arises because the
free energy profile in this situation always has a barrier whose
height scales to leading order as $N^{3/5}$, which in
turn implies exponential barrier crossing times.

In conclusion we have
presented a systematic study of the impact of crowding on
translocation dynamics. We have shown how crowding can lead to the
emergence of novel exponents characterizing the scaling of
translocation time with polymer length and crowding volume
concentration. The existence of regimes where the translocation time
depends very weakly on polymer  length, apart from being of
theoretical significance, suggests the possibility of designing
filters for a tunable range of polymer lengths and also has
implications for ``synchronized'' transport
of proteins or nucleic acids from a wide range of sizes in the cellular context.

The authors would like to thank Phil Pincus
for helpful discussions and also acknowledge support
from MRL Program of the NSF under Award number DMR00-80034 and NSF
Grant number DMR05-03347.
AG would also like to acknowledge start-up funds from UC Merced.

\end{document}